\def\rn{\noindent\parshape 2 0truecm 8.8truecm 0.3truecm 8.5truecm}
\def\nn#1 #2{#1, #2.}				
\def\nnn#1 #2 #3{#1, #2. #3.}			
\def\nnnn#1 #2 #3 #4{#1, #2. #3. #4.}		
\def\nnnnn#1 #2 #3 #4 #5{#1, #2. #3. #4. #5.}	
\def\dualand{, \&\hbox{ }}				
\def\multiand{, \&\hbox{ }}				

\def\rg#1;#2;#3;#4;#5;#6 {\par\rn#1 #2, {\it #3}, {\bf #4}, #5 (``#6'') \par}
\def\rf#1;#2;#3;#4;#5 {\par\rn#1 #2, {\it #3}, {\bf #4}, #5\par}
\def\rfbook#1;#2;#3;#4;#5 {{\frenchspacing\par\rn#1 #2, {\it #3} (#4: #5)\par}}
\def\rfproc#1;#2;#3;#4;#5;#6 {{\frenchspacing\par\rn#1 #2, in {\it #3}, ed. #4 (#5: #6)\par}}
\def\rfprep#1;#2;#3  {{\par\rn#1 #2, #3\par}}
\def\rfprepp#1;#2;#3 {{\par\rn#1 #2, #3\par}}

\def\expec#1{\langle#1\rangle}

\def\etal{{\frenchspacing\it et al.}}
\def\ie{{\frenchspacing\it i.e.}}
\def\eg{{\frenchspacing\it e.g.}}
\def\etc{{\frenchspacing\it etc.}}
\def\rms{rms}

\def\beq#1{\begin{equation}\label{#1}}
\def\eeq{\end{equation}}
\def\beqa#1{\begin{eqnarray}\label{#1}}
\def\eeqa{\end{eqnarray}}
\def\eq#1{equation~(\ref{#1})}
\def\Eq#1{Equation~(\ref{#1})}
\def\eqn#1{~(\ref{#1})}

\def\spose#1{\hbox to 0pt{#1\hss}}
\def\simlt{\mathrel{\spose{\lower 3pt\hbox{$\mathchar"218$}}
     \raise 2.0pt\hbox{$\mathchar"13C$}}}
\def\simgt{\mathrel{\spose{\lower 3pt\hbox{$\mathchar"218$}}
     \raise 2.0pt\hbox{$\mathchar"13E$}}}
\def\simpropto{\mathrel{\spose{\lower 3pt\hbox{$\mathchar"218$}}
     \raise 2.0pt\hbox{$\propto$}}}

\def\ed{\end{document}}

\def\e{{\bf e}}

\def\n{{\bf n}}
\def\r{{\bf r}}
\def\x{{\bf x}}
\def\y{{\bf y}}
\def\z{{\bf z}}
\def\xt{\tilde{\x}}
\def\yt{\tilde{\y}}
\def\bzero{{\bf 0}}
\font\bfmath=cmmib10
\def\err{\hbox{\bfmath\char'042}}	
\def\vmu{\hbox{\bfmath\char'026}}	

\def\rh{\widehat{\r}}
\def\A{{\bf A}}
\def\B{{\bf B}}
\def\C{{\bf C}}
\def\E{{\bf E}}
\def\F{{\bf F}}

\def\I{{\bf I}}
\def\M{{\bf M}}
\def\N{{\bf N}}
\def\P{{\bf P}}

\def\R{{\bf R}}
\def\SS{{\bf S}}
\def\NN{\N}
\def\Sig{{\bf\Sigma}}
\def\Nt{\tilde{\N}}
\def\St{\tilde{\SS}}
\def\Lamb{{\bf\Lambda}}
\def\W{{\bf W}}
\def\tr{\hbox{tr}\,}

\def\ith{i^{th}}
\def\jth{i^{th}}
\def\kth{i^{th}}

\def\tensormult{\otimes}

\documentstyle[emulateapj]{article}
\begin{document}


\journalid{337}{15 January 1989}
\articleid{11}{14}

\submitted{Submitted to ApJ August 31 1998; accepted February 8}

\title{COMPARING AND COMBINING CMB DATASETS}

\author{
Max Tegmark
\footnote{Institute for Advanced Study, Princeton, 
NJ 08540; max@ias.edu}$^,$\footnote{Hubble Fellow}
}

\begin{abstract}
One of the best ways of 
spotting previously undetected systematic errors in CMB experiments
is to compare two independent observations
of the same region.
We derive a set of tools for 
comparing and combining CMB data sets, applicable also
in the common case where the two have different
resolution or beam shape and therefore do not measure the same signal.
We present a consistency test that is better than a $\chi^2$-test
at detecting systematic errors.
We show how two maps of different angular resolution can be combined
without smoothing the higher resolution down to the lower one,
and generalize this to arbitrary beam configurations.
We also show how lossless foreground
removal can be performed even for foreground models involving 
scale dependence, latitude dependence and spectral index variations 
in combination.
\end{abstract}



\section{INTRODUCTION}

Undetected systematic errors are one of the main obstacles 
to using cosmic microwave
background (CMB) measurements to constrain cosmological models.
One of the best ways to address this problem is to compare experiments whose
data sets overlap both in sky coverage and angular scale, 
to see whether they are consistent.\footnote{
The best way to address the
problem is clearly to design CMB experiments to be more 
immune to systematic errors in the first place.  
The next best thing to do is carefully examine the raw time-ordered 
data from an experiment for specific forms of systematic 
errors that may be expected, and
for general signs of systematic errors for data removal or correction. 
The cross-check between experiments that are discussed in this paper
are by no means a substitute for this, but rather a way of 
catching additional systematic errors that have slipped through
the cracks and not been detected by the team that
reduced the raw data.
}
If they are consistent, a useful second step is to 
combine them into a single sky map retaining all their cosmological
information.

Both of these steps 
are simple for maps with identical resolution and beam shape. 
For {\eg} the 53 and 90 GHz COBE DMR maps (Bennett {\etal} 1996), 
comparing (1) was done by subtracting the maps and 
checking whether the difference was consistent with pure noise, whereas 
combining (2) was done by simply averaging the two maps, weighting pixels
by their inverse variance.
Unfortunately, both steps are usually more complicated. 
MAP, Planck and most current
experiments have different angular resolution in different channels. 
Many current experiments probe 
the sky in an even more complicated way, with {\eg} double beams,
triple beams, interferometric beams or complicated elongated 
software-modulated beams. Correlated noise further complicates
the problem.

Despite these difficulties, 
precision comparisons between different experiments are 
crucial. Some of the best evidence so far 
for detection of CMB fluctuations comes from  
the success of such comparisons
in the past --- between FIRS and DMR (Ganga {\etal} 1993),
Tenerife and DMR (Lineweaver {\etal} 1995), 
MSAM and Saskatoon (Knox {\etal} 1998; hereafter K98), 
two years of Python
data (Ruhl {\etal} 1995), three years of Saskatoon data 
(Tegmark {\etal} 1996a) and two flights of MSAM (Inman {\etal} 1997).
The fact that many of these non-COBE data sets were contaminated by
systematic errors made the success of these cross-checks 
even more encouraging.

Data sets are currently growing rapidly in number, size and quality,
often overlapping.
It is therefore quite timely to develop methods that generalize 
both steps (1) and (2) to arbitrary experiments.
This is 
the purpose of the present {\it Letter}.
In the larger context of CMB data analysis, this is important 
between the steps of mapmaking 
(Wright {\etal} 1996; Wright 1996; Tegmark 1997a) and 
%
%
%
%
%
%
%
%
%
power spectrum estimation (Tegmark 1997b; Bond {\etal} 1998) 
in the pipeline.

\section{Notation}

Let us first establish some notation that will be used throughout
this paper. 
Consider a pixelized CMB sky map 
at some resolution
consisting of $m$ numbers $x_1,...,x_m$, where $x_i$ is the 
temperature in the $\ith$ pixel. 
Suppose two experiments $i=1,2$ have measured 
$n_i$ numbers $y_1,...,y_{n_i}$, each probing some linear combination of the
sky temperatures $x_i$. 
Grouping these numbers into vectors $\x$, $\y_1$ and $\y_2$ of length
$m$, $n_1$ and $n_2$, 
we can generally write\footnote{
This assumes both that the experimental data
have perfect linearity, and that there are no non-zero experimental offsets.
Ideally, experimentalists should model and remove both
nonlinearities and offsets as part of their data reduction, thereby making
\eq{ModelEq1} applicable to their final data product.
If offsets of an unknown amplitude remain nonetheless, 
multiplication of a data set $\y=\A\x+\n+offsets$ by a 
matrix $\P$ that projects out
these offsets will produce a new data set $\y_1=\P\y$ that satisfies
\eq{ModelEq1}, merely with the slightly more complicated noise
vector $\n_1=\P\n$ and with $\A_1=\P\A$.
A detailed example of this procedure can be 
found in de Oliveira {\etal} (1998).
}
\beq{ModelEq1}
\y_1 = \A_1\x+\n_1,\quad
\y_2 = \A_2\x+\n_2
\eeq
for some known matrices $\A_i$ incorporating the beam shapes and
some random noise vectors $\n_i$ with zero mean $(\expec{\n_i}=\bzero)$.
We will refer to $\x$ as the ``true sky''. $\y_1$ and $\y_2$ can be 
either time-ordered data or some linear combination thereof, for instance
pixelized maps.
It is sometimes useful to define the larger matrices and vectors
\beq{GroupingEq}
\A\equiv\left(\A_1\atop\A_2\right),\quad
\y\equiv\left(\y_1\atop\y_2\right),\quad
\n\equiv\left(\n_1\atop\n_2\right),\quad
\eeq
which allows us to rewrite \eq{ModelEq1} as
\beq{ModelEq2}
\y=\A\x+\n.
\eeq
Let us write the noise covariance matrix as
\beq{NoiseCovDefEq}
\N\equiv\expec{\n\n^t}
=\left(\begin{tabular}{cc}
$\N_1$&$\N_{12}$\\
$\N_{12}^t$&$\N_2$
\end{tabular}\right),
\eeq
where 
$\N_1\equiv\expec{\n_1\n_1^t}$,
$\N_2\equiv\expec{\n_2\n_2^t}$ and
$\N_{12}\equiv\expec{\n_1\n_2^t}$.

We derive useful consistency
tests in \S 2 both for the special case of identical observations
$(\A_1=\A_2)$ and for the general case $\A_1\ne\A_2$, then show how to 
combine data sets without destroying information in \S 3. 

\section{Comparing data sets: are they consistent?}

\subsection{The ``null-buster'' test}

Let us first consider the simple case where the two data sets measure the same
thing, \ie, $\A_1=\A_2$. This often applies for two different channels 
of the same experiment at the same frequency.
We can then form a difference map $\z\equiv\x_1-\x_2$, which in the
absence of systematic errors should consist of pure noise.

Consider the null hypothesis $H_0$ that 
such a data set $\z$ consists of pure noise,
\ie, $\expec{\z}=\bzero$, $\expec{\z\z^t}=\NN$
for some noise covariance matrix $\NN$.
Suppose we have reason to suspect that
the alternative hypothesis $H_1$ is true, where
$\expec{\z}=\bzero$, $\expec{\z\z^t}=\NN+\SS$
for some signal covariance matrix $\SS$,
and want to try to rule out $H_0$ by using a 
test statistic $q$ that is a quadratic function of the
data:
\beq{qDefEq}
q\equiv\z^t\E\z^t=\tr\{\E\z\z^t\}.
\eeq
Depending on whether $H_0$ or $H_1$ is true, 
the mean of $q$ will be
$\expec{q}_0=\tr\E\NN$ or
$\expec{q}_1=\tr\E\NN + \tr\E\SS$, respectively. 
If $\z$ has a multivariate Gaussian probability distribution\footnote{
We will only make the assumption of Gaussianity for the CMB and 
the detector noise, not for the systematic errors.
In fact, systematics, such as foreground signals, data
spikes and atmospherics seldom have a 
Gaussian probability distribution, but they
vanish under the null hypothesis that we are trying to rule out.
},
then $q$ will have a variance $(\Delta q)^2=2\,\tr\E\NN\E\NN$ if 
$H_0$ is true. Therefore the quantity
$\nu\equiv(q-\expec{q}_0)/\Delta q$
gives the number of standard deviations (``sigmas'') by which
the observed $q$-value exceeds the mean expected under
the null hypothesis.
If we observe $\nu\gg 1$, we can thus conclude
that $H_0$ is ruled out at high significance.
Which choice of $\E$ has the greatest statistical power
to reject $H_0$ if $H_1$ is true, 
\ie, which $\E$ maximizes the expectation
value
\beq{nuExpecEq}
\expec{\nu}\equiv {\expec{q}_1-\expec{q}_0\over\Delta q}
= {\tr\E\SS\over\left[2\,\tr\E\NN\E\NN\right]^{1/2}}?
\eeq
Since rescaling $\E$ by a constant leaves $\expec{\nu}$ invariant,
let us for simplicity normalize $\E$ so that the denominator equals
unity. We thus want to maximize $\tr\E\SS$ subject to the
constraint that $\tr\E\NN\E\NN=1/2$.
Using the method of Lagrange multipliers with
$L=\tr\E\SS - \lambda\tr\E\NN\E\NN/2$ and differentiating $L$ with
respect to the components of $\E$, this 
gives the solution $\E\propto\NN^{-1}\SS\NN^{-1}$.
This leaves us with our optimal
``null-buster'' statistic\footnote{
K98 discuss a test using the likelihood ratio,
which is the best solution to a slightly different problem.
Translated into our notation, it corresponds to the choice 
$\E=\N^{-1}-[\N+\SS]^{-1}=\N^{-1}\SS[\N+\SS]^{-1}$.
Note that whereas \eq{NullbusterEq} is independent of 
the normalization of $\SS$ (the ``shape'' of the 
signal matters, but not its amplitude), the likelihood ratio 
test requires an assumed amplitude. 
}
\beq{NullbusterEq}
    \nu \equiv {\z^t\NN^{-1}\SS\NN^{-1}\z - \tr\NN^{-1}\SS
    \over
    \left[2\,\tr\left\{\NN^{-1}\SS\NN^{-1}\SS\right\}\right]^{1/2}},
\eeq
which will rule out the null hypothesis $H_0$ with the largest
average significance if $H_1$ is true.
For the special case $\SS=\NN$, we see that this reduces to a
standard $\chi^2$-test with $\nu=(\chi^2-n)/\sqrt{2n}$, 
$\chi^2\equiv\z^t\NN^{-1}\z$. 
Whenever we have reason to suspect systematic errors of a certain
form (producing a signal $\propto\SS$), the null-buster test will
thus be more sensitive to these systematic errors than the
$\chi^2$-test, which is a general-purpose tool.
This issue is elaborated in K98, which also provides 
a useful general discussion of consistency tests.

\bigskip
\bigskip
\centerline{\vbox{\epsfxsize=8.9cm\epsfbox{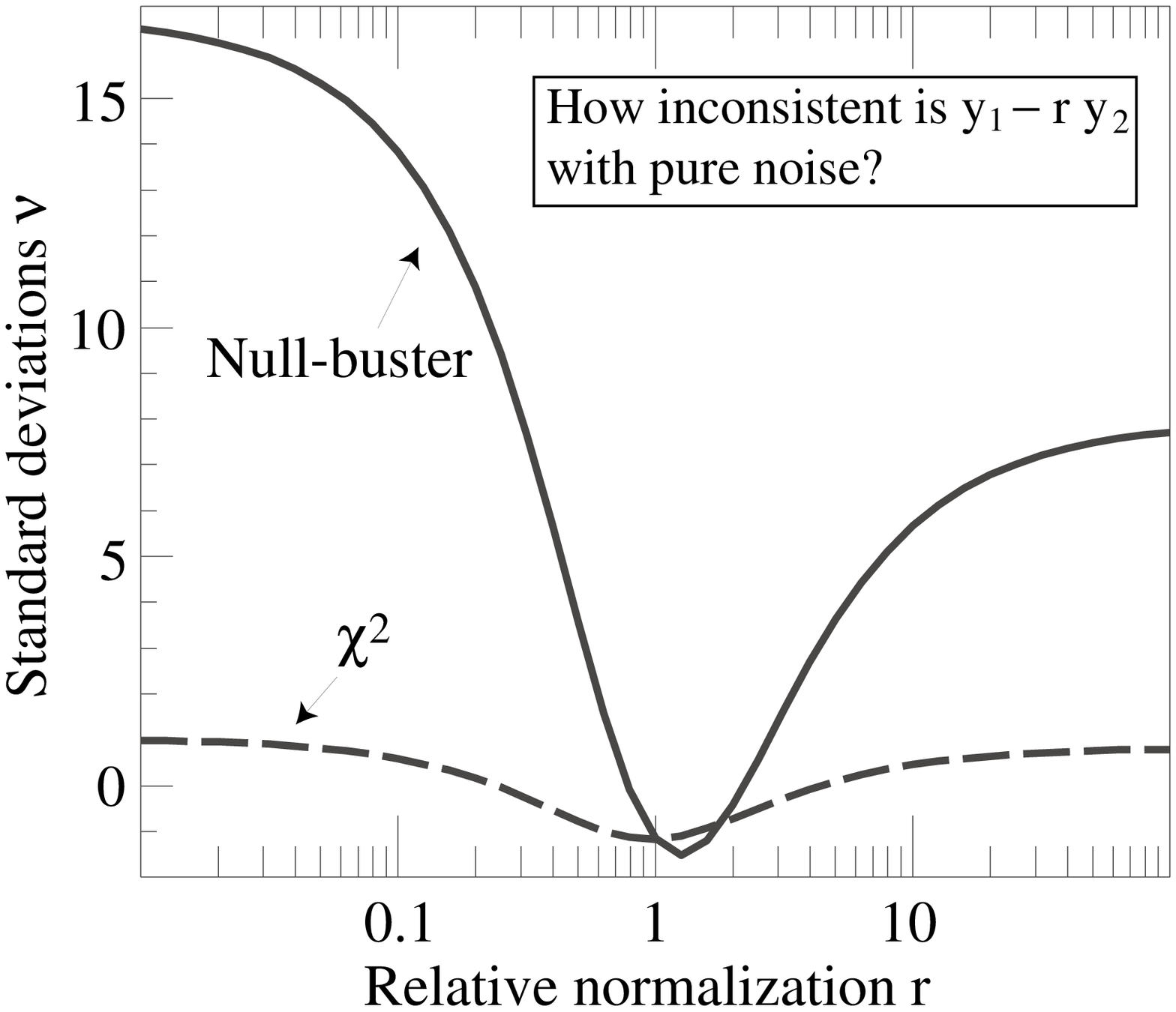}}}
\vskip-0.2cm
\figcaption{
The null-buster test against calibration errors was applied 
to the QMAP experiment
(see de Oliveira-Costa et al.~1998). The figure
(from Devlin et al.~1998) shows 
the number of $\sigma$ at which signal is detected in the 
weighted flight 1 Ka-band difference maps $\y_1-r\y_2$. 
The $\chi^2$-test is seen to be weaker. 
}
\bigskip

\subsection{An example: calibration errors}

The null-buster is useful for comparing two CMB maps $\y_1$ and 
$\y_2$ that have the same shape, beam size and pixelization.
Let $\SS$ denote the expected contribution
to ${\y_i}{\y_i}^t$ from CMB fluctuations, \ie,
$\SS=\A_1\C\A_1^t=\A_2\C\A_2^t$,
where $\C$ is the map covariance matrix
\beq{CdefEq}
\C_{ij} = \sum_{\l=2}^\infty 
\left({2\l+1\over 4\pi}\right)P_\l(\rh_i\cdot\rh_j)C_\l;
\eeq
the unit vector $\rh_i$ gives the direction towards the $\ith$ pixel
and $C_\l$ is the expected or observed CMB power spectrum
(the normalization of $C_\l$ is irrelevant here --- only the shape matters).
Now suppose one or both of the data sets $\y_1$ and $\y_2$ have 
a linear calibration error, {\ie}, are off by some constant multiplicative
factors. 
Consider a difference map of
the form $\z\equiv\y_1-r\y_2$ for some factor $r$, and plot 
$\nu$ as a function of $r$ using \eq{NullbusterEq} with
the null hypothesis being that $\z$ is pure noise, 
$\ie$, that $\NN\equiv\NN_1+r[\NN_{12}+\NN_{12}^t]+r^2\NN_2$.
An example is shown in Figure 1.
If $\nu\gg 1$ for $r=1$, then we have a significant detection
of signal not common to the two maps.
If $\nu(1)\gg 1$ but $\nu(r)\simlt 1$ for some other $r$-value,
this would show that there is a relative error of $r$ in the
normalization between the two maps. 
If the maps are at different
frequencies, this could also indicate that they are dominated
by foreground contamination with a frequency 
dependence different from the CMB.

\subsection{If the beam shape differs}

Above we found the null-buster to be a useful consistency
test when applied to $\z\equiv\y_1-r\y_2$, since $\A_1=\A_2$ implied that
$\y_1-\y_2$ should consist of mere noise and be independent of the 
(a priori unknown) signal.
If $\A_1\ne\A_2$, this is no longer true, since the two data sets are not
measuring the same thing. 
We therefore perform our null-buster test with 
the difference map redefined to be 
\beq{zRedefEq}
\z\equiv\A_3\y_1 - r \A_4\y_2
\eeq
for some matrices $\A_3$ and $\A_4$, and choose these matrices so that
the new data vectors $\A_3\y_1$ and $\A_4\y_2$ 
measure at least approximately the same sky signal.
Substituting \eq{ModelEq1} into \eq{zRedefEq}
shows that this corresponds to the requirement
\beq{SNgoalEq}
\A_3\A_1\approx\A_4\A_2,
\eeq
which would make 
$\z$ approximately signal-independent for $r=1$.
In addition to \eq{SNgoalEq}, we clearly want 
$\A_3\A_1$ and $\A_4\A_2$ to have as large rank as possible,
to avoid destroying more information than necessary
(otherwise say $\A_3=\A_4=\bzero$ would do the trick).

If one data set can be written as a linear combination of the other
(say $\A_2 = \M\A_1$ for some matrix $\M$), then the best choice
is clearly $\A_3=\M$, $\A_4=\I$.
This is the case for identically sampled maps
where $\y_1$ has higher resolution than $\y_2$, as well as for cases
where a map (say DMR or QMAP) is compared with more complicated
weighted averages (say Saskatoon) of the same sky region at the same
or lower resolution. 
This approach was adopted by, {\eg}, Lineweaver {\etal} (1995), 
for comparing the (smoothed) Tenerife data to DMR.

We will now tackle the problem for the general case. 
Our solution (there may be others) 
involves performing a signal-to-noise eigenmode analysis 
(Bond 1994; Bunn \& Sugiyama 1995; Tegmark {\etal} 1996b) three times,
in an unusual way. We will first present this procedure with no 
derivation, then show that it solves our problem.
%
%
%
%
%
We start by solving the generalized eigenvalue problems
\beqa{SNeq1}
\left[\A_1\C_1\A_1^t\right]\B_1&=&\N_1\B_1\Lamb_1,\\
\label{SNeq2}
\left[\A_2\C_2\A_2^t\right]\B_2&=&\N_2\B_2\Lamb_2,
\eeqa
where the eigenvectors are the columns of the matrices $\B_1$ and $\B_2$,
normalized so that $\B_i^t\N_i\B_i=\I$,
and the corresponding eigenvectors are elements of the 
diagonal matrices $\Lamb_1$ and $\Lamb_2$, sorted in decreasing order.
We then reduce the width of $\B_1$ and $\B_2$ by throwing away
all eigenvectors with eigenvalues below some cutoff $\lambda_{min}$,
and define a new smaller data set 
%
%
\beq{PurgedyDefEq}
\yt = \left(\B_1^t\y_1\atop\B_2^t\y_2\right).
\eeq
This will have the covariance matrix 
$\expec{\yt\yt^t} = \St + \Nt$, where
\beqa{PurgedCovEq1}
\St&=&\left(\begin{tabular}{cc}
$\Lamb_1$&$\B_1^t\A_1\C\A_2^t\B_2$\\
$\B_2^t\A_2\C\A_1^t\B_1$&$\Lamb_2$
\end{tabular}\right),\\
\Nt&=&\left(\begin{tabular}{cc}
$\I$&$\B_1^t\N_{12}\B_2$\\
$\B_2^t\N_{12}^t\B_1$&$\I$
\end{tabular}\right).
\eeqa
We then solve the generalized eigenvalue problem
\beq{SNeq3}
\St\B = \Nt\B\Lamb,
\eeq
where the eigenvectors are the columns of the matrix $\B$,
normalized so that $\B^t\Nt\B=\I$,
and the corresponding eigenvectors are elements of the 
diagonal matrix $\Lamb$, sorted in {\it increasing} order.
Finally, we reduce the width of $\B$ by throwing away
all eigenvectors with eigenvalues {\it above} some cutoff 
$\lambda_{max}$, leaving us with a matrix of the form
\beq{BdecompEq}
\B=\left(\B_3\atop\B_4\right).
\eeq
By choosing a tiny cutoff (say $\lambda_{max}=10^{-2}$),
we ensure that the transformed data vector
$\B^t\yt$ is completely dominated by 
detector noise, with a for all practical 
purposes negligible CMB signal. 
This means that
\beq{AirporterEq}
\B^t\yt=[\B_3^t\B_1^t\A_1+\B_4^t\B_2^t\A_2]\x \approx\bzero,
\eeq
so we can solve our original problem by defining
\beqa{EqualizerEq}
\A_3&\equiv&\B_3^t\B_1^t,\\ 
\A_4&\equiv&-\B_4^t\B_2^t. 
\eeqa

Why was the first eigenmode step necessary?
We went through the extra trouble of solving
equations\eqn{SNeq1} and\eqn{SNeq2} and 
applying the cutoff $\lambda_{min}$ because
otherwise, a number $(\B^t\yt)_i$
in our final data vector could be noise-dominated for two 
different reasons:
\begin{enumerate}
\itemsep-1mm
\item 
Because the signal contribution from $\y_1$ approximately 
cancels that from $\y_2$.
\item
Because it is a noise-dominated mode from 
$\y_1$ or $\y_2$ (or some combination thereof).
\end{enumerate}
It is clearly only the first case that interests us when comparing data sets. 
In the latter case, applying the null-buster 
only tests for systematic errors internally, 
within each data set, and this is best done before 
comparing it with other data sets. We therefore throw away
all noise-dominated modes from the individual data sets,
choosing say $\lambda_{min}=1$, before proceeding to
the final eigenvalue problem\eqn{SNeq3}.
A lower threshold may be appropriate as well --- as long as we
choose $\lambda_{min}\gg\lambda_{max}$, we know that 
the lack of signal in $\B^t\yt$ will be due mainly to subtracting the
data sets.

%
%

\section{Combining data sets}

\subsection{Combining maps of different beam shape}

Suppose that we have performed all the tests described above
and conclude that the data sets $\y_1$ and $\y_2$ are consistent. 
We then wish to 
simplify future calculations by combining the two data 
sets into a single map $\xt$, inverting the 
(usually over-determined) system of linear
equations\eqn{ModelEq2}.
A physically different but mathematically identical problem 
was solved in Tegmark (1997a), showing that the minimum-variance choice
\beq{ComboEq1}
\xt = [\A^t\N^{-1}\A]^{-1}\A^t\N^{-1}\y
\eeq
retains all the cosmological information of the original data sets.
Substituting this into \eq{ModelEq2} shows that
the combined map is unbiased ($\expec{\xt}=\x$) and that the 
pixel noise $\err\equiv\xt-\x$ has the covariance matrix
\beq{ComboCovarEq}
\Sig\equiv\expec{\err\err^t}=[\A^t\N^{-1}\A]^{-1}.
\eeq
A common special case is that where the two data sets have 
uncorrelated noise $(\N_{12}=\bzero)$,  simplifying the solution to 
\beqa{ComboEq2}
\xt&=&\Sig\left[\A_1^t\N_1^{-1}\y_1 + \A_2^t\N_2^{-1}\y_2\right],\\
\Sig&=&[\A_1^t\N_1^{-1}\A_1+\A_2^t\N_2^{-1}\A_2]^{-1}.
\eeqa
An even simpler case occurs if the first data set is already a sky map 
(say the QMAP map), so that $\A_1=\I$, and we wish to combine it with
a more complicated data set covering the same sky region
(say the Saskatoon observations). For this case,  
\eq{ComboEq2} can be rewritten as
\beq{ComboEq3}
\xt=\y_1 + \Sig\A_2^t\N_2^{-1}(\y_2-\A_2\y_1),
\eeq
which has a simple interpretation. The vector $\A_2\y_1$ is just
map 1 convolved with the observing strategy of experiment 2,
so the factor $(\y_2-\A_2\y_1)$ contains only noise. 
The map $\xt$ is thus obtained by correcting $\y_1$ with a pure noise term 
that partially cancels some of its noisiest modes.

This important case applies also to combining two maps
with different angular resolution.
For instance, if $\y_1$ and $\y_2$ have narrow Gaussian beams
of resolution $\theta_1$ and $\theta_2$, with $\theta_2>\theta_1$, then
we define $\x$ to be the true sky map smoothed by $\theta$ and 
have
$(\A_2)_{ij} = \exp[-\theta_{ij}^2/2\theta^2]/2\pi\theta^2$
where $\theta_{ij}=\cos^{-1}(\rh_i\cdot\rh_j)$ is the 
angular separation between pixels $i$ and $j$ and 
$\theta=(\theta_2^2-\theta_1^2)^{1/2}$ is the extra smoothing in map 2.
Despite occasional claims to the contrary, this shows that
two maps at different resolution can be combined without 
destroying any information, without first degrading
the higher resolution down to the lower one by smoothing.
Instead, \eq{ComboEq1} will use the lower resolution map $\y_2$ to 
improve the accuracy of the large-scale fluctuations
in $\y_1$
(as was done by Schlegel {\etal} 1998 when combining the 
DIRBE and IRAS maps), 
retaining all the information from both maps.

Note that even the case of two identical 
maps ($\A_1=\A_2=\I$) can be non-trivial.
\Eq{ComboEq2} shows that the optimal combination is
$\xt=\Sig\left[\N_1^{-1}\y_1 + \N_2^{-1}\y_2\right]$.
This was used in the combined QMAP analysis
(de Oliveira-Costa {\etal} 1998), and 
reduces to separate averaging for each pixel
only if the two maps have vanishing or identical noise correlations.

\subsection{Combining maps at different frequencies to remove foregrounds}

As described in Tegmark (1998) and further elaborated in White (1998),
foregrounds can be treated as simply an additional source of noise that is 
correlated between channels $(\N_{12}\ne\bzero)$. 
This means that 
if we have $d$ data sets measured at different frequencies $\nu_\alpha$,
$\alpha=1,2,...,d$, each defined by their own
matrix $\A_\alpha$, the best\footnote{
Specifically, under the assumption that 
foregrounds and detector noise have
multivariate Gaussian probability distributions, one 
one can show (Tegmark 1997) that this foreground
removal method retains all the cosmological information present in the
multifrequency set of input maps, i.e., constitutes an
information-theoretically lossless form of data compression
reducing all data down to a single CMB map.
One does not need to assume that the CMB itself is Gaussian.
Systematics such as foreground signals, data
spikes, and atmospherics seldom have a Gaussian probability distribution
-- in this more general case, 
the removal method is no longer strictly lossless,
but retains the feature that it minimizes the 
total rms of foregrounds and noise 
assuming only their finite second moment.
} 
way to combine them is still given by
\eq{ComboEq1} --- we simply need to include more physics in 
the noise covariance matrix $\N$.
Let us be more explicit about this.
The noise covariance matrix $\N$ will be of size
$n\times n$, where $n=\sum_{\alpha=1}^d n_\alpha$,
\ie, the total number of numbers in all data sets combined.
We will therefore write the elements of $\N$ as
$\N_{\alpha i\beta j}$, where $\alpha$ and $\beta$ determine 
the data sets and $i$ and $j$ the numbers therein.
If there are
$f$ foreground components, this noise matrix 
becomes a sum $\N=\sum_{k=0}^f\N^{(k)}$,
where $\N^{(0)}$ is the contribution from instrumental noise
and the other terms are the contributions from 
foregrounds (synchrotron emission, bremsstrahlung, 
dust, point sources, {\etc}).
Each of these foreground matrices will be of the form
\beq{ForegNeq1}
\N^{(k)}_{\alpha i\beta j} = 
\F^{(k)}_{\alpha\beta}[\A_\alpha\C^{(k)}\A_\beta^t]_{ij},
\eeq
where the $d\times d$ matrix $\F^{(k)}$ gives the 
covariance of the $\kth$ foreground between frequencies, and 
the $m\times m$ matrix $\C^{(k)}$ gives its spatial covariance
between pixels. For example, if the data sets are identical maps 
($\A_\alpha=\I$) at $d=2$ different frequencies, we obtain the 
$(2m)\times(2m)$ block matrix
\beq{ForegNeq2}
\N^{(k)}_{\alpha i\beta j} = 
\F^{(k)}\tensormult\C^{(k)} = 
\left(\begin{tabular}{cc}
$F^{(k)}_{11}\C^{(k)}$&$F^{(k)}_{12}\C^{(k)}$\\
$F^{(k)}_{21}\C^{(k)}$&$F^{(k)}_{22}\C^{(k)}$
\end{tabular}\right).
\eeq
Explicit models specifying the foreground dependence 
on frequency and position (the matrices $\F$ and $\C$)
can be found in Tegmark (1998). 
Although it has been common to characterize $\C$ by a power spectrum
as in \eq{CdefEq}, the above formalism clearly works even if we break 
the isotropy by introducing an additional 
dependence on galactic latitude, say.

This shows how to best combine data sets at different frequencies.
When combining different multifrequency experiments, it is desirable 
to first combine corresponding maps at the same frequency and apply the
null-buster test for systematic errors. The different frequencies can
then be merged afterwards, as a second step.

\section{CONCLUSIONS}

We have derived a set of useful tools for 
comparing and combining CMB data sets, all based on simple
matrix operations,
and drawn the following conclusions:
\begin{enumerate}
\itemsep-1mm
\item The ``null-buster'' test is better at detecting 
systematic errors than a simple $\chi^2$-test.

\item Such a consistency test can be performed even between two
experiments with quite different beam shape and observing strategy.

\item When combining two maps of different angular resolution, 
one need not smooth the higher resolution down to the lower one.

\item When combining two identical maps, one should generally not
do the averaging separately for each pixel.

\item Our map merging method also handles the case of {\eg} Planck,
where the beams are elliptical rather than round and the different
detectors have different beam orientations.
  
\item The foreground removal method of Tegmark (1998) is a special case
of the combination technique we derived, and can be carried out 
even for foreground models involving frequency dependence, 
scale dependence, latitude dependence and spectral index variations 
in combination.

\end{enumerate}
As the available CMB data sets continue to increase in quantity and quality, 
it will be useful to perform such cross-checks against systematic errors
and incrementally combine all consistent data sets 
into a single state-of-the art map containing our
entire knowledge of the CMB sky.\footnote{
The signal and noise covariance matrices $\SS$ and $\N$
of such a map
could of course be rather complicated. However, 
the same complication is encountered 
if one maintains separate experiment-specific maps and only attempts to 
combine the power spectrum measurements. This is because 
the sample variance of the band power measurements from 
different experiments are generally not independent, and
computation of the correlation requires knowledge of 
these matrices. A disadvantage of combining only at the power spectrum 
level is that this generally sacrifices useful information on
relative phases,
whereas the merged map retains all the cosmological 
information from both data sets.
For instance, near degeneracies in the noise covariance
matrices of two maps can often be broken by combining them. 
} 

Both theories and new observations can then be tested
against this combined map as it gradually 
grows in size, quality and resolution.

\smallskip
The author wishes to thank Angelica de Oliveira-Costa and Martin White
for helpful comments.
Support for this work was provided by
NASA though grant NAG5-6034 and 
Hubble Fellowship HF-01084.01-96A from STScI, operated by AURA, Inc. 
under NASA contract NAS5-26555. 


\bigskip
\bigskip
\bigskip
This paper is available at
{\it h t t p://www.sns.ias.edu/$\tilde{~}$max/comparing.html}

\end{document}